\documentclass[11pt]{article}
   \usepackage{latexsym}
   \usepackage{amsmath}
   \usepackage{amssymb}
   \font\scten=cmitt10

\title{Finding best possible constant for a polynomial inequality}

\author{{\sc Lu Yang}\/$^{1}$\,\,\quad {\sc Ju Zhang}\/$^{2}$\\
  $^{1}${\small Chengdu Institute of Computer Applications}\\
  {\small CAS, Chengdu 610041, China}\\
  $^{2}${\small Chongqing Institute of Green and Intelligent Technology}\\
  {\small CAS, Chongqing 400714, China}\\
 {\tt luyang@casit.ac.cn\,\,\quad zhangju@cigit.ac.cn}}

\date{}

\begin{document}

\maketitle

 Given a multi-variant polynomial inequality with a parameter, how to find
 the best possible value of this parameter that satisfies the inequality?
 For instance, find the greatest number $ k $ that satisfies
 $ a^3+b^3+c^3+ k(a^2b+b^2c+c^2a)-(k+1)(ab^2+bc^2+ca^2)\geq 0 $
 for all nonnegative real numbers $ a,b,c $.
 Analogues problems often appeared in studies of inequalities and were dealt
 with by various methods. In this paper, a general algorithm is proposed for
 finding the required best possible constant. The algorithm can be easily
 implemented by computer algebra tools such as Maple.

\section{Construct a set including all critical values}
 \mbox{\hspace{8pt}} This paper is aimed at the kind of problems as the following:
 \medskip

 {\bf Problem 1.}\, Given a polynomial $F(k,\, x,\, y,\, z)$,
 find the greatest number $k$ that satisfies $F \geq 0$ for all real numbers $x,\, y,\, z$.
 \medskip

 Constraint $F\geq 0$ define a closed set, so the greatest or least value of $k$ can only be reached on the boundary $F(k,\, x,\, y,\, z)=0$.

 Therefore, the required greatest number must be a critical value of $k$ where $k$ is taken as an implicit function defined by $F(k,\, x,\, y,\, z)=0$.

\medskip

  Based on the above consideration, the suggested procedure of solving Problem 1 consists of two steps:
 \begin{itemize}
 \item construct a real number set $S$ which includes all the critical values of $k$, where $k$ is taken as an implicit function defined by $F=0$
 \item distinguish the greatest or least value of $k$ from other members of $S$
 \end {itemize}
 \medskip

 In the present paper, the treatment method for the first step is as same as that in the author's earlier articles \cite{Yang00,Yang01,Yang08}, however, for the second step, the approach is taken in a somewhat different way - just use a {\small MAPLE} internal package without any external software such as {\small BOTTEMA}.
\medskip

 {\bf Example 1.}\ \ Find the greatest number $ k $ that satisfies
 \begin{equation}
    a^3+b^3+c^3+ k\,(a^2b+b^2c+c^2a)-(k+1)(ab^2+bc^2+ca^2)\geq 0
 \end{equation}
 for all nonnegative real numbers $ a,b,c $.
 
  By $f$ denote $ a^3+b^3+c^3+ k\,(a^2b+b^2c+c^2a)-(k+1)(ab^2+bc^2+ca^2) $.
 \medskip
 
 To confirm the existance of such a greatest value, we need to prove that the feasible set of $k$  has an upper bound.
 There are several ways to be used, say, substituting $2,3,1$ for $a,b,c$\/ respectively, the inequality (1) becomes $11-2\,k\geq 0$, hence $k<6$. So, such a greatest value actually exists since the feasible set is a closed one.
 \medskip

   In order to convert the problem to an unconstrained optimization, we replace $a,b,c$\/ with $x^2,y^2,z^2$ respectively, and denote the resulted polynomial by $F$,
   $$F:= x^6+y^6+z^6+k\,(x^4y^2+y^4z^2+z^4x^2)-(k+1)(x^2y^4+y^2z^4+z^2x^4).$$

  According to the procedure stated above, first construct a set including all the critical values of $k$. To do this, we may employ an elimination algorithm, {\it Successive Resultant Projection}\/, which is sketched as follows:
 \medskip

  Given a polynomial $\varphi$ in
  $x_1,\,x_2,\,\cdots$, compute the resultant of $\varphi$ and $\frac{\partial{\varphi}}{\,\,\partial{x_1}}$ with
   respect to $x_1$, remove the multiple factors, and denote that by $\varphi_1$; compute the resultant of $\varphi_1$ and $\frac{\partial{\varphi_1}}{\partial{x_2}}$ with
   respect to $x_2$, remove the multiple factors, and denote by $\varphi_2$; $\cdots$,
    repeat this procedure successively until the last resultant becomes a univariate polynomial.
  \medskip

  We may use the following self-compiled short program with Maple to remove multiple factors:

 \noindent\verb|> powerfree:= proc (poly, vset) local P, W, fs;|\\
 \verb|    P:= 1; fs:= factors(poly)[2];|\\
 \verb|    if nops(fs)= 1 then return fs[1][1] end if;|\\
 \verb|    for W in op(map(L->L[1], fs)) do|\\
 \verb|       if has(W, vset) = true then P:= P*W end if|\\
 \verb|    end do;|\\
 \verb|    P|\\
 \verb|  end proc|

 \noindent This program is able to remove not only multiple factors but other useless or redundant factors as well.

 \medskip

 Let's go back to Example 1.

 At first, when $k$ is taken as an implicit function defined by $f(k,\, a,\, b,\, c)=0$, find a set including all the critical values of $k$\/ by means of the Successive Resultant Projection, with Maple.

 \noindent{\small\verb|> F:= x^6+y^6+z^6+k*(x^4*y^2+x^2*z^4+y^4*z^2)-(k+1)*(x^4*z^2+x^2*y^4+y^2*z^4)|}\\
\verb|> f1:= powerfree(resultant(F, diff(F, x), x), k)|\\
\verb|> f2:= powerfree(resultant(f1, diff(f1, y), y), k)|

\noindent Now we have
\begin{equation}
f_2 = (k^2+k+1)(k^4+2\,k^3-\frac{107}{7}\,k^2-\frac{114}{7}\,k-\frac{89}{7})(k^4+2\,k^3-5\,k^2-6\,k-23)
(k^2+k+\frac{19}{27})
\end{equation}
Perform the operation of real root isolating,

\noindent\verb|> realroot(f2)|

\noindent that resulted in a list of intervals:
\begin{equation}
\bigg{[}\bigg{[}-\frac{597}{128}, -\frac{149}{32}\bigg{]}, \bigg{[}-\frac{447}{128}, -\frac{223}{64}\bigg{]}, \bigg{[}\frac{159}{64}, \frac{319}{128}\bigg{]}, \bigg{[}\frac{117}{32}, \frac{469}{128}\bigg{]}\bigg{]}
\end{equation}

\noindent whereof each interval contains one and only one real root of $f_2$.
\medskip

So, $k$ has at most 4 critical values because the real roots of $f_2$ include all the critical values of $k$, consequently, if the greatest value of $k$ (denoted by $k_{\rm max}$) exists, it must belong to one of the 4 intervals above.
\medskip

On the other hand, there may also be some intervals where none critical value of $k$ is in, because resultant computation can sometimes produce extra-factors.
\medskip

So far the first step of our procedure has been done.

\section{Directly down to Real-Root-Classification}

Consider a semi-algebraic system:
 \begin{equation*}\left\{\begin{array}{l}
\varPhi(k)=0,\\
\varPsi(k,x,y,z)\geq 0, \\
k-\alpha >0,\, \beta -k >0,
\end{array}\right.\end{equation*}
where $\alpha,\,\beta$\/ are constants.

We ask for a necessary and sufficient condition on parameters $x,\,y,\,z$\/ such that the equation $\varPhi(k)=0$ has a specified number of roots satisfying all the inequalities above.
\vspace{1pt}

More specifically, following the denotations in Section 1, let
{\small\begin{eqnarray*}
f_2=&&\mbox{\hspace{-12pt}}(k^2+k+1)(k^4+2\,k^3-\frac{107}{7}\,k^2-\frac{114}{7}\,k-\frac{89}{7})(k^4+2\,k^3-5\,k^2-6\,k-23)
(k^2+k+\frac{19}{27})\\
  F=&&\mbox{\hspace{-12pt}}x^6+y^6+z^6+k\,(x^4y^2+y^4z^2+z^4x^2)-(k+1)(x^2y^4+y^2z^4+z^2x^4)
\end{eqnarray*}}
and ask for a necessary and sufficient condition on parameters $x,\,y,\,z$\/ such that the equation $f_2=0$ has a unique real root satisfying all the inequalities in the following system:
 \begin{equation}\left\{\begin{array}{l}
f_2=0,\\\vspace{1\jot}
F\geq 0,\\
\vspace{1\jot}
k-\frac{159}{64} >0,\, \frac{319}{128} -k >0.
\end{array}\right.\end{equation}

There has been a Maple function, {\it RealRootClassification}\/, which is specially designed to solve this kind of problems. To get the required condition, we first start Maple and load some relative internal packages as follows.

\noindent\verb|> with(RegularChains):|\\
\verb|> with(ParametricSystemTools):|\\
\verb|> with(SemiAlgebraicSetTools):|

\noindent Then define an order of the variables:

\noindent\verb|> R:= PolynomialRing([k,x,y,z]):|

\noindent and choose a more direct output

\noindent\verb|> infolevel[RegularChains]:= 1:|

\noindent Input the relative polynomials:

\noindent{\small\verb|> f2:= (k^2+k+1)*(k^4+2*k^3-107/7*k^2-114/7*k-89/7)*(k^4+2*k^3-5*k^2-6*k-23)|}

\mbox{\hspace{20pt}}{\small\verb|*(k^2+k+19/27)|}

\noindent{\small\verb|> F:= x^6+y^6+z^6+k*(x^4*y^2+x^2*z^4+y^4*z^2)-(k+1)*(x^4*z^2+x^2*y^4+y^2*z^4)|}

\noindent Now, by calling

\noindent
\verb|> RealRootClassification([f2],[F],[k-159/64,319/128-k],[ ],3,1..n,R)|

\noindent the screen displays the following output:
\smallskip

   {\tt There is always given number of real solution(s)!}
 \smallskip

 \noindent That means, for any reals $x,\,y,\,z$\/, polynomial $f_2$ always has some root in $\big{(}\frac{159}{64},\, \frac{319}{128}\big{)}$ which satisfies $F\geq0$. Such a root is unique because $f_2$ has only one root in this interval.
 By $k_1$  denote this unique root, then
  \begin{equation}
   x^6+y^6+z^6+k_1(x^4y^2+y^4z^2+z^4x^2)-(k_1+1)(x^2y^4+y^2z^4+z^2x^4)\geq 0
 \end{equation}
 for all reals $ x,y,z $. Hence,
  \begin{equation}
   a^3+b^3+c^3+ k_1(a^2b+b^2c+c^2a)-(k_1+1)(ab^2+bc^2+ca^2)\geq 0
 \end{equation}
 for all nonnegative real numbers $ a,b,c $.

 On the other hand, by $k_0$ denote the greatest value of $k$ taken over $\big{[}\frac{159}{64},\, \frac{319}{128}\big{]}$ that satisfying $F\geq 0$ for all $x,\,y,\,z$. Since $k_0$ is a local maximum, so is a critical value of $k$, hence a real root of $f_2$.

 However, $f_2$ has only one root in $\big{[}\frac{159}{64},\, \frac{319}{128}\big{]}$, so $k_0=k_1$, that is, $k_1$ is the local maximum of $k$ taken over $\big{[}\frac{159}{64},\, \frac{319}{128}\big{]}$.
\medskip

  To compute $k_1$, by calling

 \noindent\verb|> k1 = fsolve(f2, k, 159/64..319/128)|

 \noindent the screen displays the floating-point value of $k_1$:
   $$ k_1 = 2.484435332 \cdots$$

 In this way, for any interval in list (3), say, $\big{[}\frac{117}{32}, \frac{469}{128}\big{]}$, we can decide if it contains a local maximum of $k$ or not.


 By calling

 \noindent
 \verb|> RealRootClassification([f2],[F],[k-117/32,469/128-k],[ ],3,1..n,R)|

\noindent the screen displays the following output:

\verb|The system has given number of real solution(s) IF AND ONLY IF|

\verb|[[R[1]<=0]|

\verb|OR|

\verb|[0<=R[1] 0<R[2]]|

 {\boldmath $\cdots \cdots \cdots \cdots$}

  That means, the root of $f_2$ in $\big{[}\frac{117}{32}, \frac{469}{128}\big{]}$ does not satisfy $F\geq 0$ when $R_1>0,\, R_2<0$, i.e. it even cannot satisfy $F\geq 0$ for some $x,\, y,\, z$, not to mention being a local maximum of $k$ we asked for.
 \medskip

 Note that the intervals of list (3) are displayed automatically in increasing order, i.e. the numbers in one interval are not greater than those in the next interval. Therefore, on  distinguishing which interval contains the global maximum of $k$, the first two intervals can be ignored, and on the other hand, the last interval, $\big{[}\frac{117}{32}, \frac{469}{128}\big{]}$, has already been excluded, so $k_1$ is not only a local maximum but also the global maximum of $k$, the greatest number which satisfies $F\geq 0$ for all real numbers $x,\, y,\, z$.
 \medskip

 By establishing the corresponding relation between each interval of list (3) and the real roots of polynomial $f_2$, we know that $k_1$ is the unique real root of $k^4+2\,k^3-5\,k^2-6\,k-23$ in {\small$\big{[}\frac{159}{64},\, \frac{319}{128}\big{]}$}. Then,
 $$k_1 = \, -\frac{1}{2}+\frac{1}{2}\,\sqrt{13+16\,\sqrt{2}}\, \approx\, 2.484435332 \cdots $$

 \section{Alternate: convert to inequality proving}

 Next, we consider a analogous problem.

 {\bf Example 2.}\ \ Find the greatest number $ k $ that satisfies
 \begin{equation}
  2\,(a^3+b^3+c^3)+ 3\,k\,a\,b\,c-(k+2)(a^2b+b^2c+c^2a)\geq 0
 \end{equation}
 for all nonnegative real numbers $ a,b,c $.

 By $f$ denote $2\,(a^3+b^3+c^3)+ 3\,k\,a\,b\,c-(k+2)(a^2b+b^2c+c^2a)$.
 \medskip

  Observe that $f$ is a decreasing function with respect to $k$\/, because
 \begin{equation}
  \frac{\partial{f}}{\partial{k}}=3\,abc-(a^2b+b^2c+c^2a)\leq 0
 \end{equation}
 by Arithmetic-mean-Geometric-mean inequality.
 \medskip

   In order to convert the problem to an unconstrained optimization, we replace $a,b,c$\/ with $x^2,y^2,z^2$ respectively, and denote the resulted polynomial by $F$,
   $$F:= 2\,(x^6+y^6+z^6)+3\,k\,x^2y^2z^2-(k+2)(x^4y^2+y^4z^2+z^4x^2).$$
 \indent  Perform the first step of the procedure as done previously, to construct a set including all the critical values of $k$\/ by means of the Successive Resultant Projection:

\noindent\verb|> F := 2*(x^6+y^6+z^6)+3*k*x^2*y^2*z^2-(k+2)*(x^4*y^2+y^4*z^2+z^4*x^2)|\\
\verb|> f1:= powerfree(resultant(F, diff(F, x), x), k)|\\
\verb|> f2:= powerfree(resultant(f1, diff(f1, y), y), k)|

\noindent Now we have

$f_2:= \big{(}k^4-\frac{38}{3}\,k^3+8\,k^2-\frac{16}{3}\big{)}(k^3+6\,k^2+12\,k-46)
(k^3+42\,k^2+264\,k+152)(k+2)(k-4)\\{}\\
\mbox{\hspace{40pt}}\big{(}k^4+\frac{112}{25}\,k^3-\frac{1224}{25}\,k^2+
\frac{1472}{25}\,k+\frac{1424}{25}\big{)}.$
\medskip

Perform the operation of real root isolating,

\noindent\verb|> realroot(f2)|

\noindent that resulted in a list of intervals:
{\small\begin{eqnarray}
&&\mbox{\hspace{-6pt}}\bigg{[}\bigg{[}-\frac{4413}{128}, -\frac{1103}{32}\bigg{]}, \bigg{[}-\frac{1273}{128}, -\frac{159}{16}\bigg{]}, \bigg{[}-\frac{883}{128}, -\frac{441}{64}\bigg{]}, [-2, -2], \bigg{[}-\frac{41}{64}, -\frac{81}{128}\bigg{]}, \nonumber \\
&&\bigg{[}-\frac{81}{128}, -\frac{5}{8}\bigg{]},\bigg{[}-\frac{75}{128}, -\frac{37}{64}\bigg{]}, \bigg{[}\frac{227}{128}, \frac{57}{32}\bigg{]}, \bigg{[}\frac{347}{128}, \frac{87}{32}\bigg{]}, \bigg{[}\frac{27}{8}, \frac{433}{128}\bigg{]}, [4, 4], \bigg{[}12, \frac{1537}{128}\bigg{]}\bigg{]}.
\end{eqnarray}}
\noindent where each interval contains one and only one real root of $f_2$. Among the 12 intervals, we need to decide which one contains the global maximum of $k$.
\medskip

 It was pointed out that $f$ is a decreasing function with respect to $k$\/, so is $F$.  This fact allows us to use a more efficient method.

  Let $[\alpha,\, \beta]$ be an interval and $\alpha$ is not a critical value of $k$\/. We have the following assertion:
 \medskip

  {\bf Proposition 1.}\, If $F(\alpha,\, x,\, y,\, z)\geq 0$ holds for all real numbers $x,y,z$\/, but $F(\beta,\, x,\, y,\, z)\geq 0$ does not hold for some reals $x,y,z$\/, then, the global maximum of $k$ must belong to $[\alpha,\, \beta]$.
  \medskip

  This is almost obvious. By $k_{\rm max}$ denote the global maximum of $k$.  That $k_{\rm max}>\alpha$ because $k_{\rm max}$ is ``the greatest number'' satisfying $F\geq 0$ for all reals $x,y,z$\/, and $k_{max}\ne \alpha$ for $\alpha$ is not a critical value of $k$\/.\,\, On the other hand, $F$ is decreasing with respect to $k$, since $F(\beta,\, x,\, y,\, z)\geq 0$ does not hold for some reals $x,y,z$\/, not to mention any number greater than $\beta$, so we have $\alpha < k_{\rm max} <\beta$\/.
  \medskip

  Now, the problem of finding $k_{\rm max}$ has been converted to inequality proving.
 \medskip

  For instance, to check if the interval $\big{[}\frac{227}{128}, \frac{57}{32}\big{]}$ contains $k_{\rm max}$, introduce a slack variable $t$, and make use of {\small RealRootClassification} as follows.
 \medskip

\noindent
\verb|> F:= 2*(x^6+y^6+z^6)+3*k*x^2*y^2*z^2-(k+2)*(x^4*y^2+y^4*z^2+z^4*x^2)|\\
\verb|> g1:= subs(k=227/128, F)|\\
\verb|> g2:= subs(k=57/32, F)|
\smallskip

\noindent i.e. where $g_1,\, g_2$ denote $F(\frac{227}{128},\, x,\, y,\, z),\, F(\frac{57}{32},\, x,\, y,\, z)$ respectively.

\noindent \verb|> with(RegularChains):|\\
\verb|> with(ParametricSystemTools):|\\
\verb|> with(SemiAlgebraicSetTools):|\\
\verb|> R:= PolynomialRing([t,x,y,z]):|\\
\verb|> infolevel[RegularChains]:= 1:|
\smallskip

\noindent and then, by calling
\smallskip

\noindent
\verb|> RealRootClassification([g1+t],[ ],[t],[ ],3,0,R)|

\noindent the screen displays the following output:

\verb|There is always given number of real solution(s)!|
 \smallskip

 Let's interpret what the meaning is. The first argument of the input, $g_1+t$, means given an equation $g_1+t=0$, where $t$ is a slack variable; the third argument, $t$, means a requirement $t>0$; the sixth argument, $0$, means ``none real root''. So, the input means: {\scten regarding $t$ as the unknown and $x,y,z$ as parameters, to find a sufficient and necessary condition under which the equation $g_1+t=0$ has none positive root.}
 \medskip

 The output, {\tt there is always given number of real solution(s)}, gives the answer that the equation $g_1+t=0$ has none positive root for any $x,y,z$, i.e. $g_1\geq 0$ for all reals $x,y,z$.
\medskip

Let's go on, by calling
\smallskip

\noindent
\verb|> RealRootClassification([g2+t],[ ],[t],[ ],3,0,R)|

\noindent the screen displays the following output:

\verb|The system has given number of real solution(s) IF AND ONLY IF|

\verb|[0 < R[1]]|

\verb|where|

\verb|R[1] = 64*x^6-121*x^4*y^2+171*x^2*y^2*z^2-121*x^2*z^4+ |{\boldmath $\cdots \cdots $}
\smallskip

\noindent That means, $g_2+t=0$ has none positive root only if $R_1>0$ but it will have a positive root when $R_1<0$, i.e. $g_2\geq 0$ does not hold for some reals $x,y,z$\/.

 Therefore, according to Proposition 1, the global maximum $k_{\rm max}$ must belong to the interval {\small $\big{[}\frac{227}{128}, \frac{57}{32}\big{]}$}.
\medskip

  To compute $k_{\rm max}$, by calling

 \noindent\verb|> k[max] = fsolve(f2, k, 227/128..57/32)|

 \noindent the screen displays the floating-point value of $k_1$:
   $$ k_{\rm max} = 1.779763150 \cdots$$

   By establishing the corresponding relation between each interval in list (9) and the real roots of polynomial $f_2$, we know that $k_{\rm max}$ is the unique real root of $k^3+6\,k^2+12\,k-46$. Then,
 $$ k_{\rm max} = \, 3\,\sqrt[3]{2}-2\, \approx\, 1.779763150 \cdots $$

 In Example 1 and Example 2, the best possible constant can be written in radicals, but frequently cannot be in some other cases.
\medskip

 {\bf Remark.}\ \ List (9) contains a ``degenerate'' interval $[\alpha,\, \alpha]$ indicates that $\alpha$ is a rational root of $f_2$. Then, $\alpha$ is the global maximum of $k$\/ if and only if $F(\alpha,\, x,\, y,\, z)\geq 0$ for all $x,y,z$\/ and any number which is greater than $\alpha$\/ does not satisfy this constraint.
 The same argument also applies to the list (3) in Example 1.
\medskip

 {\bf Example 3.}\cite{Franklin80}\ \ Find the greatest number $k$ that satisfies
 \begin{equation}
  a^2b^4-k\,ab^3+\sqrt{a^2+b^4}\,ab^3 +b^3+a\geq 0
 \end{equation}
 for all nonnegative real numbers $ a,b $.

 By $f$ denote $a^2b^4-k\,ab^3+\sqrt{a^2+b^4}\,ab^3 +b^3+a$.

  Clearly $f$ is a decreasing function with respect to $k$\/.

 \medskip

   In order to convert the problem to an unconstrained optimization, we replace $a,b,c$\/ with $x^2,y^2,z^2$ respectively, and denote the resulted polynomial by $\varPhi$,
   $$\varPhi := x^4y^8-k\,x^2y^6+\sqrt{x^4+y^8}\,x^2y^6+y^6+x^2.$$
   \indent To remove the radical, let $h:= x^4+y^8-u^2$ so that $\varPhi$ can be replaced with the rational polynomial,
 \begin{equation}
   F:= x^4y^8-k\,x^2y^6+u\,x^2y^6+y^6+x^2.
 \end{equation}

 Perform the first step of the procedure as done previously, to construct a set including all the critical values of $k$\/ by means of the Successive Resultant Projection:

\noindent\verb|> F := x^4*y^8-k*x^2*y^6+u*x^2*y^6+y^6+x^2|\\
\verb|> h := x^4+y^8-u^2|\\
\verb|> f0:= powerfree(resultant(F, h, u), k)|\\
\verb|> f1:= powerfree(resultant(f0, diff(f0, x), x), k)|\\
\verb|> f2:= powerfree(resultant(f1, diff(f1, y), y), k)|

\noindent The screen displays that $f_2$ is the product of three factors which are of degrees 34, 32, 8, respectively. The factor of degree 34 is:
{\small\begin{equation}
\mbox{\hspace{-16pt}}k^{34}-\frac{929}{729}\,k^{32}-\frac{22}{27}\,k^{31}-
\frac{5086745}{78732}\,k^{30}+
\frac{5923}{2187}\,k^{29}+\frac{10698803575}{136048896}\,k^{28}-\cdots
\end{equation}}
Perform the operation of real root isolating,

\noindent\verb|> realroot(f2)|

\noindent that resulted in a list of intervals:
{\small\begin{eqnarray}
&&\mbox{\hspace{-32pt}}\bigg{[}\bigg{[}-\frac{5}{8}, -\frac{79}{128}\bigg{]}, \bigg{[}\frac{17}{128}, \frac{69}{512}\bigg{]}, \bigg{[}\frac{73}{256}, \frac{37}{128}\bigg{]}, \bigg{[}\frac{51}{64}, \frac{103}{128}\bigg{]},
\bigg{[}\frac{147}{128}, \frac{37}{32}\bigg{]},\bigg{[}\frac{69}{16}, \frac{553}{128}\bigg{]} \bigg{]}.
\end{eqnarray}}
\noindent where each interval contains one and only one real root of $f_2$. Among the 6 intervals, we need to decide which one contains the global maximum of $k$.
\medskip

  To check if the interval $\big{[}\frac{69}{16}, \frac{553}{128}\big{]}$ contains $k_{\rm max}$, we make use of {\small RealRootClassification } as follows.
 \smallskip

\noindent
\verb|> g1:= subs(k=69/16, F)|\\
\verb|> g2:= subs(k=553/128, F)|
\smallskip

\noindent i.e. $g_1:=x^4y^8-\frac{69}{16}\,x^2y^6+u\,x^2y^6+y^6+x^2$,

 \mbox{\hspace{8pt}}$g_2:=x^4y^8-\frac{553}{128}\,x^2y^6+u\,x^2y^6+y^6+x^2$.

\noindent \verb|> with(RegularChains):|\\
\verb|> with(ParametricSystemTools):|\\
\verb|> with(SemiAlgebraicSetTools):|\\
\verb|> R:= PolynomialRing([u,x,y]):|\\
\verb|> infolevel[RegularChains]:= 1:|
\smallskip

\noindent and then, by calling
\smallskip

\noindent
\verb|> RealRootClassification([h],[u],[-g1],[ ],2,0,R)|

\noindent the screen displays the following output:

\verb|There is always given number of real solution(s)!|

\noindent That means, regarding $u$ as the unknown, the equation $h=0$ has none nonnegative roots satisfying $g_1<0$ for any $x,\, y$\/, i.e. when
$u=\sqrt{x^4+y^8}$,\, we have $g_1\geq 0$ for all real numbers $x,\, y$\/.
\medskip

Let's go on, by calling
\smallskip

\noindent
\verb|> RealRootClassification([h],[u],[-g2],[ ],2,0,R)|

\noindent the screen displays the following output:

\verb|The system has given number of real solution(s) IF AND ONLY IF|

\verb|[0<= R[1]  0<R[2]]|

\verb|OR|

\verb|[R[2]<0]|

{\boldmath $\cdots \cdots $}
\smallskip

\noindent That means, regarding $u$ as the unknown, when $R_1<0$ and $R_2>0$, the equation $h=0$ will have a positive root such that $g_2<0$, i.e. $g_2\geq 0$ does not hold for some reals $x,y$\/.

 Therefore, according to Proposition 1, the global maximum $k_{\rm max}$ must belong to the interval {\small $\big{[}\frac{69}{16}, \frac{553}{128}\big{]}$}.
\medskip

  To compute $k_{\rm max}$, by calling

 \noindent\verb|> k[max] = fsolve(f2, k, 69/16..533/128)|

 \noindent the screen displays the floating-point value of $k_1$:
   $$ k_{\rm max} = 4.315351626 \cdots$$

   By establishing the corresponding relation between each interval in list (13) and the real roots of polynomial $f_2$, we know that $k_{\rm max}$ is the greatest real root of an irreducible polynomial of degree 34.
   \medskip

   The method used in this section does not apply to the Example 1, because the Proposition 1 is invalid as the polynomial $a^3+b^3+c^3+ k\,(a^2b+b^2c+c^2a)-(k+1)(ab^2+bc^2+ca^2) $ is neither decreasing nor increasing  with respect to $k$ for all nonnegative numbers $a,b,c$.

 \section{Conclusion}

 \mbox{\hspace{6pt}} We demonstrated a symbolic method with Maple for finding the best possible value of a parameter satisfying some constraints of inequalities and equalities. This method does not employ the external packages.
 \medskip

   A recent article \cite{Han16} gave a proof of the correctness of successive  resultant projection and proposed a simplified projection operator of which the projection scale is smaller so it is more effective for many problems. However, the optimization method provided in that article does not seem to solve some constrained problems such as the third example in the present paper.
 \medskip

   The equivalence between successive resultant projection and Brown-McCallum's projection \cite{Brown01,McCallum98,McCallum88} was proven in \cite{Yao14}.

 \bibliographystyle{abbrv}

\end{document}